\documentclass[journal]{IEEEtran}
\usepackage{gensymb}

\usepackage{cite}

\ifCLASSINFOpdf
   \usepackage[pdftex]{graphicx}
   \graphicspath{{../pdf/}{../jpeg/}}
   
   \DeclareGraphicsExtensions{.pdf,.jpeg,.png}
\else
\fi
\usepackage{amsmath}
\usepackage{array}

\ifCLASSOPTIONcompsoc
  \usepackage[caption=false,font=normalsize,labelfont=sf,textfont=sf]{subfig}
\else
  \usepackage[caption=false,font=footnotesize]{subfig}
\fi
\usepackage{fixltx2e}

\usepackage{stfloats}
\usepackage{siunitx}
\usepackage[dvipsnames,svgnames,x11names]{xcolor}
\usepackage[markup=underlined]{changes}
\definechangesauthor[color=BrickRed]{AVN}

\begin{document}
%

%
%
%

\title{Repeated Acoustic Vaporization of Perfluorohexane Nanodroplets for Contrast-Enhanced Ultrasound Imaging}

\author{Austin~Van~Namen,
        Sidhartha~Jandhyala,
        Tomas~Jordan,
        and~Geoffrey~Luke 
\thanks{A. Van Namen, S. Jandhyala, T. Jordan and G. Luke are with the Thayer School of Engineering at Dartmouth College, Hanover, NH, 03755}%
\thanks{Please send correspondence to e-mail: geoffrey.p.luke@dartmouth.edu}%
}

%


\maketitle

\begin{abstract}
Superheated perfluorocarbon nanodroplets are emerging ultrasound imaging contrast agents boasting biocompatible components, unique phase-change dynamics, and therapeutic loading capabilities. Upon exposure to a sufficiently high intensity pulse of acoustic energy, the nanodroplet’s perfluorocarbon core undergoes a liquid-to-gas phase change and becomes an echogenic microbubble, providing ultrasound contrast. The controllable activation leads to high-contrast images, while the small size of the nanodroplets promotes longer circulation times and better \textit{in-vivo} stability. One drawback, however, is that the nanodroplets can only be vaporized a single time, limiting their versatility. Recently, we and others have addressed this issue by using a perfluorohexane core, which has a boiling point above body temperature. Thus after vaporization, the microbubbles recondense back into their stable nanodroplet form. Previous work with perfluorohexane nanodroplets relied on optical activation via pulsed laser absorption of an encapsulated dye. This strategy limits the imaging depth and temporal resolution of the method. In this study we overcome these limitations by demonstrating acoustic droplet vaporization with 1.1-MHz high-intensity focused ultrasound. A short-duration, high-amplitude pulse of focused ultrasound provides a sufficiently strong peak negative pressure to initiate vaporization. When using a custom imaging sequence with a high-frequency transducer, the repeated acoustic activation of perfluorohexane nanodroplets can be visualized in polyacrylamide tissue-mimicking phantoms. We demonstrate detection of hundreds of vaporization events from individual nanodroplets with activation thresholds well below the tissue cavitation limit. Overall, this approach has the potential to result in reliable contrast-enhanced ultrasound imaging at clinically relevant depths.
\end{abstract}

\IEEEpeerreviewmaketitle

\section{Introduction}
\IEEEPARstart{G}{aseous} microbubbles augment the diagnostic and therapeutic potential of ultrasound imaging for a wide variety of applications, such as vascular mapping, lithotripsy, targeted drug delivery, tissue ablation, and clot disruption \cite{Gessner2010AdvancesUltrasound,Ferrara2007UltrasoundDelivery,Sboros2010TheImaging,Stride2010CavitationTherapy,McDannold2013NonthermalFunction,Hitchcock2010Ultrasound-assistedBubbles,Klibanov2006MicrobubbleApplications,Bader2016Sonothrombolysis}. The large acoustic impedance mismatch between the gaseous core of the microbubbles and the water-based tissue background produces ultrasound contrast. Contrast can be further enhanced by harnessing the nonlinear acoustic response of microbubbles with novel ultrasound image acquisition strategies, such as harmonic imaging or pulse inversion\cite{Simpson1999PulseAgents,Shi2000UltrasonicMicrobubbles,Quaia2007MicrobubbleUpdate}. The typical size of a microbubble contrast agent is between 1-10 \si\micro m in diameter which restricts these contrast agents to the vasculature and leads to fast clearance from the body. However, contrast agents with a nanometer-scale diameter allow for extravasation into the leaky, disorganized vasculature of tumors. Extravasation of contrast agents enhances ultrasound’s potential for molecular tumor-specific targets\cite{Campbell2008TumorNanopharmaceuticals}.

Sub-micron phase change particles, such as perfluorocarbon nanodroplets (PFCnDs), were created in a push to decrease contrast agent size in order to promote tumor extravasation and enhance \textit{in-vivo} stability\cite{Sheeran2017MethodsUltrasonography}. PFCnDs consist of a liquid perfluorocarbon core and a stabilizing shell. In their innate liquid nanodroplet state, the PFCnDs do not provide ultrasound contrast; however, they can be optically or acoustically activated to undergo a liquid to gas phase change and form micron-sized bubbles. Optical activation is produced through absorption of laser energy by a photoabsorber embedded in the nanodroplet. Acoustic activation, referred to as acoustic droplet vaporization (ADV), is achieved by exposing the PFCnDs to a negative acoustic pressure below a threshold amplitude, initiating the phase change. A sufficient peak negative pressure (PNP) allows the HnDs to vaporize by overcoming both the PFC vaporization pressure and the Laplace pressure exerted by the lipid shell on the PFC liquid core. The unique dynamics of PFCnDs have enabled their use in a wide range of applications including super-resolution imaging, disruption of the blood-brain barrier, vascular imaging, and targeted drug delivery\cite{Luke2016Super-ResolutionNanodroplets,Hallam2019TowardNanodroplets,Sheeran2016Phase-ChangeTherapy,Cao2018DrugUltrasound,Sheeran2013TowardPrinciple}.  Typically a low-boiling-point perfluorocarbon (e.g., perfluorobutane or perfluoropentane with a boiling point of -2 \degree C or 28 \degree C, respectively) is used so that the activation energy triggers irreversible vaporization to form gaseous microbubbles. Thus, PFCnDs are activatable, nano-sized contrast agents that have the potential for improved biodistribution properties\cite{Lea-Banks2019Ultrasound-responsiveReview,Loskutova2019ReviewTherapeutics,Kripfgans2000AcousticApplications,Rojas2019VaporizationRatio,Shpak2014AcousticFocusing.}.

\begin{figure}[h]
\includegraphics[scale=.5]{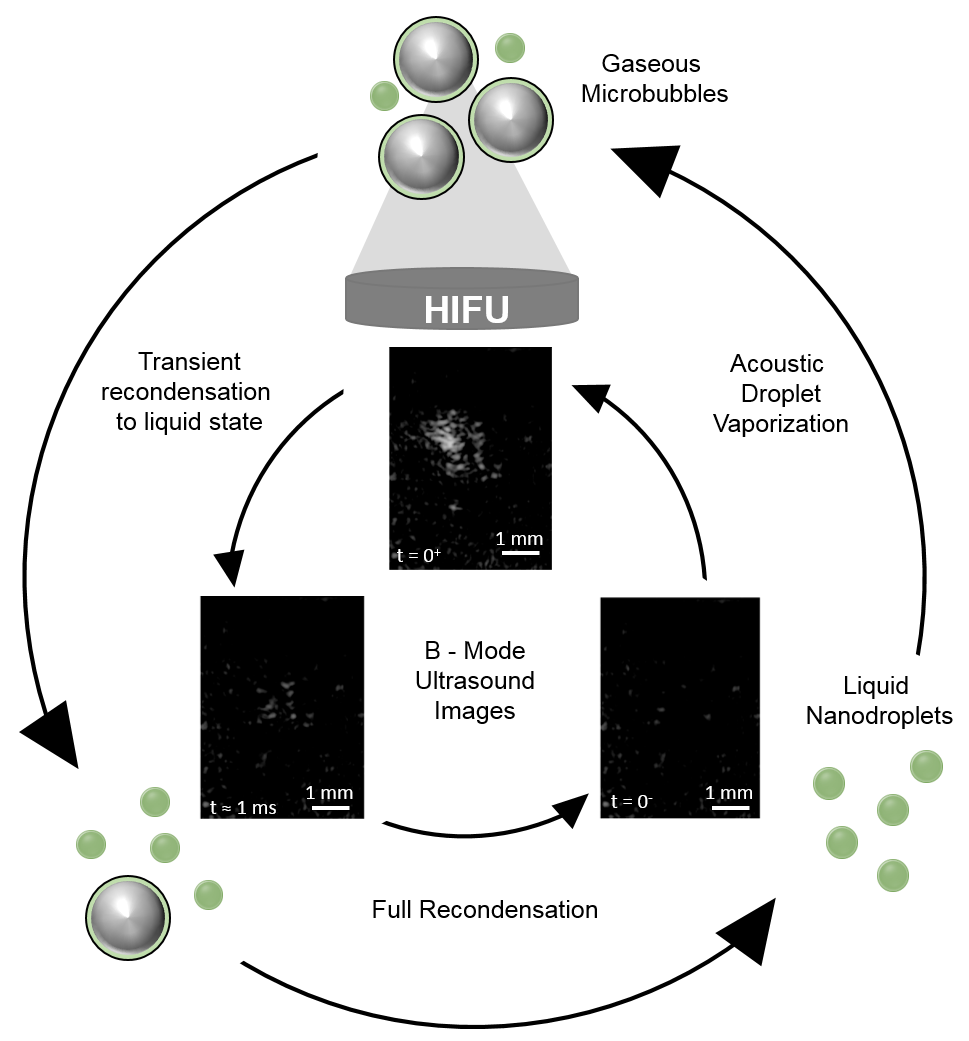}
\centering

\caption{HnD vaporization and recondensation process. Due to the high boiling point of the perfluorohexane, the gaseous microbubble returns to its nanodroplet form on the order of milliseconds.}
\label{fig:process}
\end{figure}

Although low-boiling-point PFCnDs are a promising alternative to microbubbles, additional benefits can be achieved by using a core with a higher boiling point. We and others have proposed using perfluorohexane as the core of the PFCnDs \cite{Luke2016Super-ResolutionNanodroplets,Hannah2016BlinkingImaging.,Fabiilli2009TheVaporization,Burgess2019ControlNanoemulsionsb,Strohm2012AcousticEmulsions,Ishijima2016TheNano-droplets,Dayton2006ApplicationTherapy}. Perfluorohexane has a longer carbon chain length which leads to a boiling point of 56 \degree C. The higher boiling point results in two related phenomena. First, the activation threshold is higher; more energy is needed to vaporize the particles. Second, because the boiling point is above that of the surrounding tissue, the perfluorohexane nanodroplets (HnDs) recondense back to their stable liquid nanodroplet form a short time after each vaporization event \cite{Hannah2016BlinkingImaging.,Asami2012RepeatableImaging,Hallam2018Laser-activatedOpening,Ishijima2016TheNano-droplets,Zhu2019LeveragingUltrasound}. Thus, the vaporization can be repeated.  An illustration and example of this phenomenon is depicted in Figure \ref{fig:process}.

We have previously shown that, when an optically absorbing dye is encapsulated in the particles, a laser pulse can reliably initiate the vaporization event and image processing techniques can be applied to the images to extract high contrast images of the blinking HnDs or pinpoint their location with micrometer-scale precision\cite{Hannah2016BlinkingImaging.,Luke2016Super-ResolutionNanodroplets}. Although laser vaporization of the HnDs can result in reliable and repeatable activation of the contrast agents, the method has limitations in how it can be applied. Optical absorption and scattering in tissue of near-infrared wavelength light limit the depth of HnD vaporization. Previous work shows diminished HnD activity after 5 mm for HnDs with near-infrared absorbing dyes\cite{Luke2016Super-ResolutionNanodroplets}. Additionally, optical activation is limited by the pulse repetition rate of the laser (typically 10-20 Hz). While an acoustic activation strategy could overcome these limitations, ADV of HnDs remains a challenge. The relatively high vaporization threshold makes it difficult to vaporize the HnDs with an imaging transducer.

High-intensity focused ultrasound (HIFU) is currently used in the clinic for primarily therapeutic applications\cite{Kennedy2005High-intensityTumours,Moyer2015High-intensityMicrobubbles,Hill1995HighTreatment}. It has been shown to deliver both mechanical energy and localized heating in a contained volume centimeters deep in tissue. If high-amplitude, short pulses are applied, then inertial cavitation can occur and the resulting microbubbles can be used to mechanically disrupt the tissue, a technique known as lithotripsy or histotripsy\cite{Stride2010CavitationTherapy,Hitchcock2010Ultrasound-assistedBubbles,YoshizawaHighMicrobubbles}. These techniques have been shown to result in negligible deposition of thermal energy\cite{Khokhlova2015HistotripsyApplications,Hoogenboom2015MechanicalEffects}. Conversely, if excitation pulses with a longer duty cycle are used, then the tissue can be thermally ablated. Beyond mechanical and thermal ablation, focused ultrasound has been applied at lower pressure levels for applications including neuromodulation and physical therapy\cite{Darrow2019FocusedNeuromodulation,Baker2001AEffects}.

The approach detailed in this paper harnesses HIFU energy to combine the benefits of repeatedly vaporized HnDs with the advantages of ADV. The applied HIFU pulses used for HnD vaporization have a low duty cycle (to avoid heating)\cite{Khokhlova2015HistotripsyApplications,Hoogenboom2015MechanicalEffects} and a PNP below the cavitation threshold (to avoid mechanical damage) \cite{Xu2018AcousticUltrasound,Maxwell2013ProbabilityMaterials.,Lea-Banks2019Ultrasound-responsiveReview}. High-intensity focused ultrasound is commonly applied at depths of several centimeters, thereby expanding the potential applications of HnDs beyond what can be achieved with an optical stimulus\cite{Kennedy2005High-intensityTumours,Moyer2015High-intensityMicrobubbles,Hill1995HighTreatment}. In addition, HIFU can safely be applied at kHz pulse repetition rates\cite{Miller2012OverviewConsiderations}. Faster acquisition of the HnD images would enable improved detection sensitivity through averaging, better suppression of tissue motion, and visualization of faster biological processes, such as blood flow. We have developed methods and imaging sequences to clearly visualize HnDs undergoing repeated ADV. We have conducted studies to identify the HIFU thresholds for ADV of HnDs, characterize the resulting ultrasound imaging signal, and investigate the stability of the HnDs when undergoing multiple vaporization-recondensation cycles. Overall, by using HIFU as the ADV stimulus, HnDs could be more broadly applied to the diverse and growing range of applications which utilize phase-change nanodroplets.

\section{Materials and Methods}
\subsection{Nanodroplet Fabrication}
Perfluorocarbon nanodroplets were fabricated using methods adapted from a previously published thin film hydration method \cite{Kumar2008DirectionalMoieties,Hannah2013IndocyanineImaging}. A lipid monolayer shell was comprised of 1,2-dipalmitoyl-sn-glycero-3-phosphocholine (16:0 PC (DPPC), 25 mg/mL in chloroform, Product Code: 850355C, Avanti Polar Lipids Inc, Alabaster, AL, USA) and 1,2- distearoyl-sn-glycero-3-phosphoethanolamine-N- [amino(polyethylene glycol)-2000] (DSPE-PEG(2000), 25 mg/mL in chloroform, Product Code: 880128C,  Avanti Polar Lipids Inc, Alabaster, AL, USA) in a volume ratio of 90:10 DPPC to DSPE-PEG. 200 \si\micro L of the lipid mixture and 4 mL of chloroform (99.8\% chloroform ACS reagent stabilized with ethanol, CAS 67-66-3, Oakwood Chemical Inc, Estill, SC, USA). The flask was then connected to a rotary evaporator (Heidolph, Schwabach, Germany) and spun at 50 rpm under vacuum at 250 mbar in a 37.5 \degree C water bath. This was done until the chloroform evaporated completely, leaving a lipid cake in the flask. The lipid cake was then resuspended in 4 ml of phosphate-buffered saline (PBS 1x, 21-040-CV, Corning, Corning, NY, USA) using a water bath sonicator (Symphony, VWR, Radnor, PA, USA). The lipid solution was then transferred to a scintillation vial and 250 \si\micro L of liquid perfluorocarbon was added. Either perfluoropentane (APF-29M (n-Dodecafluoropentane, 99.0\%), Fluoromed, Round Rock, TX, USA), perfluorohexane (APF-60M (Perfluorohexanes, 99.0\%), Fluoromed, Round Rock, TX, USA) or perfluorooctyl bromide (Perfluorooctyl bromide, 99\%, ACROS Organics, Fair Lawn, NJ, USA) were used to make PnDs, HnDs and OBnDs respectively. The lipid and perfluorocarbon solution was then sonicated using a microtip sonicator (Q700 with a 3.2 mm tip, QSonica, Newton, CT, USA) in an ice bath to generate PFCnDs. The total sonication energy supplied to the solution was 10 J.  The PFCnDs were centrifuged (Mini-spin, Cat. No. 022620100, Eppendorf, Hamburg, Germany) and washed twice at 4293 x g RCF for 1 minute to remove excess reagents in the supernatant. Finally, the nanodroplet pellet was resuspended in 5 ml of 50:50 mixture of deionized (DI) water and PBS. 

\subsection{Nanodroplet Characterization}
The nanodroplets were characterized using a Zetasizer Nano ZS (Malvern Instruments Inc., Westborough, MA, USA). The three types of perfluorocarbon nanodroplets used in this study are perfluoropentane (PnDs), perfluorohexane (HnDs) and perfluorooctyl bromide (OBnDs). The nanodroplets were diluted 1:100 in DI Water. Three measurements with 12 recordings each were averaged together for each sample vial. A total of three different sample vials were measured per type of nanodroplet. The sample vials were measured at 25 \degree C. The concentrations were determined by the Zetasizer Nano software. The as-prepared PFCnDs were diluted a further 100x during phantom preparation.

\subsection{Phantom Fabrication}
Both the tissue-mimicking phantoms and the HIFU transduction cone in this study were constructed using a polyacrylamide (PA) hydrogel. PA phantoms have been shown to be both optically and acoustically transparent compared to other tissue mimicking substances. The PA phantoms are well suited for HIFU studies and were selected because of their low acoustic attenuation and tunable elasticity\cite{Lafon2005GelDosimetry}. This study did not supplement the phantom with acoustic or optical scatterers or absorbers in order to more clearly visualize and investigate HnD behavior. Previous studies, however, have shown PA phantoms can closely mimic many properties of tissue \cite{Lafon2005GelDosimetry,Choi2013AUltrasound,Asami2010AcousticGels}. Several published nanodroplet studies have used PA phantoms as a suitable phantom to study vaporization\cite{Asami2010AcousticGels,Song2019FacileAgents,Hannah2014PhotoacousticNm,Phillips2013Phase-shiftHeating,Hallam2018Laser-activatedOpening}.

Phantom preparation was conducted at room temperature inside a fume hood. The polyacrylamide hydrogel was fabricated in a glass beaker on a magnetic stir plate operating at 100 rpm. To make 50 ml of the hydrogel, 33.3 ml of polyacrylamide solution (acrylamide/bis-acrylamide 29:1 30\%, A3574, Sigma, St. Louis, MO, USA) and 500 \si\micro L of 10\% w/v aqueous ammonium persulfate (A3678, Sigma, St. Louis, MO, USA) were added to 16.7 ml deionized water. The solution was then degassed in a water bath sonicator for 30 seconds to remove trapped bubbles in the phantom. 50 ul of TEMED (N,N,N’,N’ - tetramethyl - ethylenediamine, Sigma, St. Louis, MO, USA) was added as a crosslinking catalyst. When making the PFCnD phantoms, the deionized water contained 500 ul of PFCnD from the stock solution. The resulting PFCnD concentration in the phantom is therefore 100x diluted from the stock. To make the phantoms, 50 mL of crosslinked PFCnD phantom solution was pipetted into a 3D-printed (Lulzbot Mini, Lulzbot, ND, USA) 7 cm in length, 3 cm in width, and 2.5 cm in height phantom mold. The long rectangular shape was chosen for the portion of experiments where multiple imaging areas were required on a single phantom such as those shown in Figure 5.  The polyacrylamide hydrogel was then set for 10 minutes in a -80 \degree C freezer to counteract the exothermic effects of the polymerization from affecting the PFCnDs.

\begin{figure}[h]
\includegraphics[scale=.42]{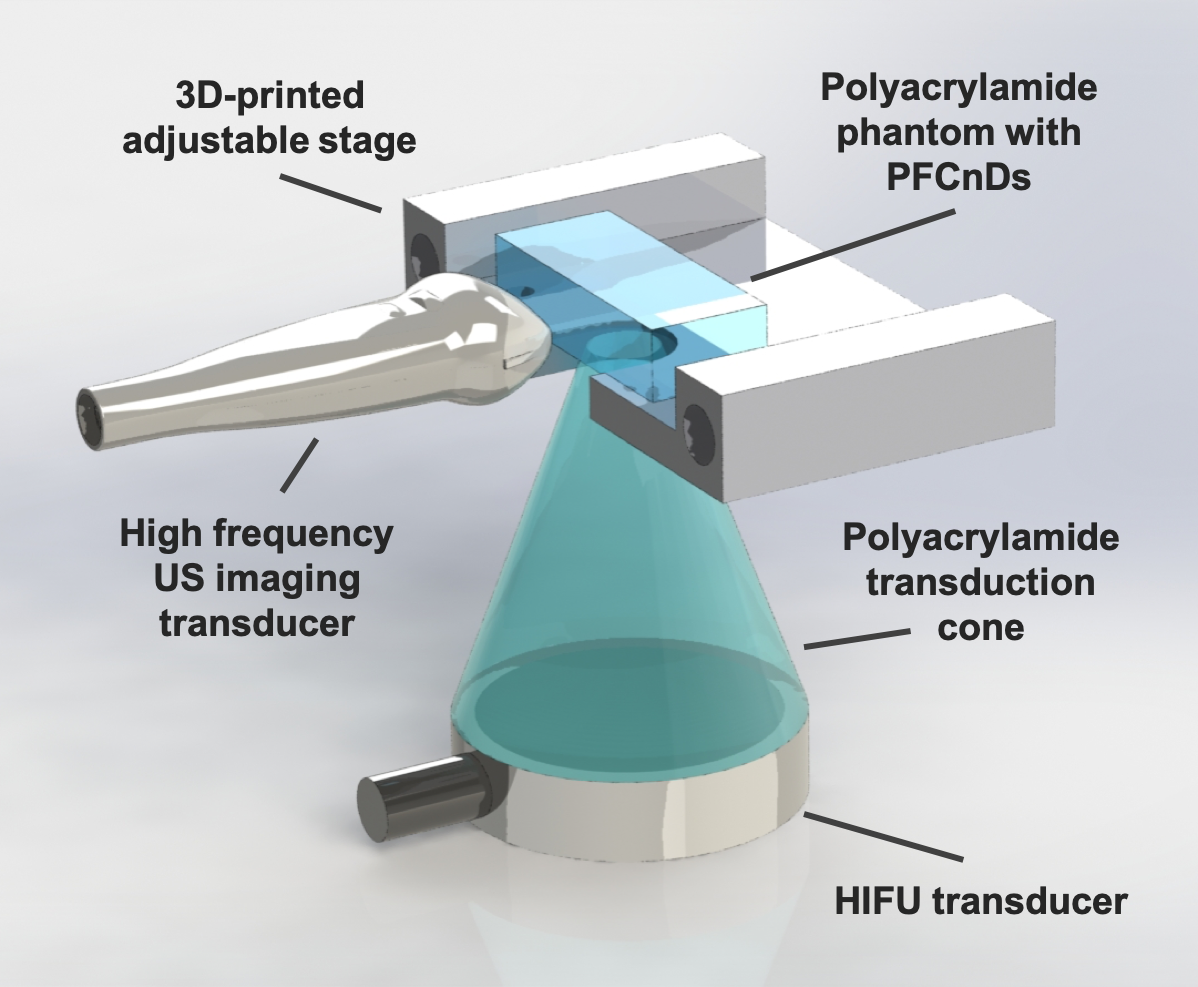}
\centering

\caption{Schematic representation of the experimental set up. A Verasonics Vantage 256 ultrasound system visualizes phase change activity of PFCnDs in tissue mimicking PA phantoms with a 15-MHz center frequency linear array transducer. The focused ultrasound pulse is conducted through a 9 cm PA hydrogel cone. The entire process is staged and aligned using 3D printed parts.
}
\label{fig:setup}
\end{figure}

For the HIFU transduction cone, 200 ml of non-PFCnD phantom solution was pipetted into a 3D printed mold to create the cone shown in Figure \ref{fig:setup}. The 3D printed mold was created using the geometry of the HIFU transducer. The resulting conical phantom is a 9-cm tall truncated cone with a base radius of 4 cm that tapers to a 0.5 cm radius at the peak. The bottom of the cone phantom is rounded to match the HIFU transducer curvature in order to provide solid contact for ultrasound transduction. A custom-designed, 3D printed apparatus precisely aligned the HIFU system and cone to achieve a consistent focal spot in the ultrasound images as shown in Figure \ref{fig:setup}. This focal spot occurs at 10 cm from the HIFU transducer which correlates to 1 cm into the PFCnD phantom.

\subsection{Ultrasound Imaging}
\begin{figure}[h]
\includegraphics[scale=.4]{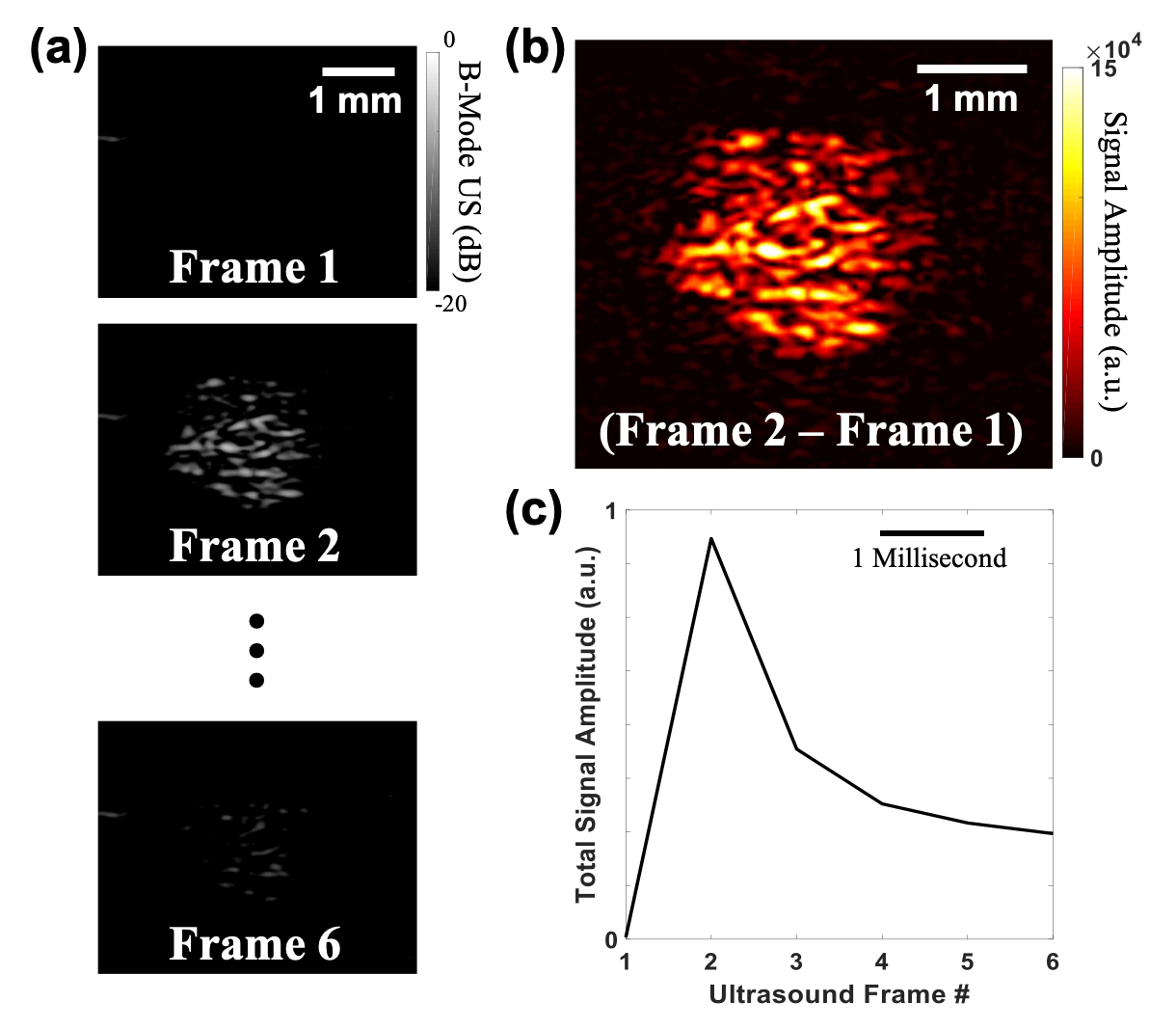}
\centering

\caption{Illustration of image visualization and processing. (a) Ultrasound B-mode  image frames of the nanodroplets, shown in grayscale, are collected via the Verasonics high frequency transducer. One frame is collected pre-HIFU and five frames post-HIFU. (b) The difference between post-HIFU and pre-HIFU frames highlights the nanodroplet vaporization and is shown in figures with a red to black colormap. (c) B-mode images are processed by taking the sum of pixel amplitudes in a square containing the HIFU focal area and condensed into a frame by frame graph. This graph helps illustrate vaporization and recondensation of HnDs.}
\label{fig:vandp}
\end{figure}

 A Verasonics Vantage 256 ultrasound imaging system was used to acquire ultrasound images and synchronize the triggering of the single-element HIFU transducer (H-151, Sonic Concepts, Bothell, WA, USA) powered by a radiofrequency power amplifier (1020L RF amplifier, E \& I, Rochester, NY, USA) connected to an impedance matching circuit. The images were acquired with a 256-element 15-MHz capacitive micromachined linear array transducer (L22-8v, Verasonics, Kirkland, WA, USA) enabling high resolution (ca. 100 \si\micro m) images of a single 2-D plane. The imaging transducer and the phantom were coupled using ultrasound gel. The Verasonics hardware control and image processing were performed using Matlab (Mathworks, Natick, MA, USA). Each B-mode ultrasound image was acquired using a plane-wave transmission from 5 steering angles equally spaced between -18 and 18 degrees. The plane wave utilized the center 128 transducer elements as transmit and receive channels for each acquisition. A 38-\si\micro s delay separated each steering angle acquisition. Representative B-mode ultrasound images of vaporization are shown in Figure \ref{fig:vandp}(a). Ultrasound frames were acquired at a rate of 2 kHz. This timing was heuristically chosen to capture HnD vaporization and recondensation kinetics. The beamforming and compounding were performed by Verasonics reconstruction algorithms.  Due to the image reconstruction time, data processing was handled asynchronously with respect to acquisition. 
 
 \subsection{Activation of PFCnDs}
 The ultrasound image acquisition was synchronized with the HIFU by a digital trigger signal output from the Verasonics system. For the HIFU setup, the trigger was received by an arbitrary waveform generator (Tektronix, Beaverton, OR, USA), which created a 10 cycle, 1.1-MHz sinusoid burst. The waveform was then amplified by a 200-W radio frequency power amplifier (1020L, E \& I). The peak negative pressure (PNP) emitted from the HIFU was controlled by the amplitude of the waveform from the function generator. Peak negative pressure measurements resulting from the power amplifier were calibrated using a needle hydrophone (Model HNA, ONDA, Sunnyvale, CA, USA). The HIFU activation occurred immediately after the first B-mode ultrasound image. A 500- \si\micro s delay between frames ensured the HIFU would not be detected in the ultrasound frame immediately following the HIFU pulse.

 \subsection{Image Processing}
 The visualization and image processing is illustrated in Figure \ref{fig:vandp}. Six B-mode ultrasound frames were acquired with the first frame occurring before the HIFU pulse. The ultrasound image acquired before HIFU was used to determine the background signal, while the ultrasound images acquired post-HIFU were used to visualize the vaporization and recondensation processes. A differential of frames, shown in Figure \ref{fig:vandp}(b), between pre-HIFU ultrasound and the first frame after HIFU exposure was used to isolate the signal from the vaporization of the HnDs. A similar differential of frames between the first and last frame post-HIFU can be used to visualize the recondensation process of HnDs. Due to the nature of the HnDs, both metrics, vaporization and recondensation, yield similar information. Hereafter, images with the red, yellow and black colormap are from the vaporization differential (i.e., the second ultrasound frame minus the first). The ultrasound signal of the nanodroplets was quantified over a 3.94x3.94 mm square focal spot, corresponding to a 150x100-pixel region in the image. The total signal amplitude is defined as the sum of all pixel values in the square focal area. This dynamic ultrasound signal is directly related to the number of microbubbles in focal area at each point in time. Tracking the focal area over consecutive frames, shown in Figure \ref{fig:vandp}(c), the signal amplitude exhibits a large increase due to PFCnD vaporization and a characteristic exponential decay associated with recondensation of the HnDs.
 
\subsection{Phantom Experiments}
Three sets of phantom experiments were performed. The first set of experiments examined the behavior of PFCnDs with different perfluorocarbon cores. The second mapped out the temperature-dependent vaporization thresholds of HnDs. The third set explored the limits of repeated vaporization of HnDs. Phantoms were prepared with nanodroplets as previously described.

For the first set of experiments five phantoms were prepared for each of four groups: PFCnD-free (blank) phantoms, PnD phantoms, HnD phantoms, and OBnD phantoms. The phantoms without PFCnDs acted as a negative control and allowed for the visualization of possible unwanted cavitation. The PnD phantoms served to visualize the behavior of PFCnDs with only a single vaporization event. Perfluoropentane is a close analog to perfluorohexane, but its shorter carbon chain length results in a lower boiling point. Any lower boiling point perfluorocarbon nanodroplet could have been chosen to achieve the same result showing single vaporization. The OBnD phantoms allowed for the investigation of PFCnDs with a much higher boiling point (and thus activation threshold). Each phantom was imaged three times at room temperature with the previously described activation sequence to visualize repeated vaporization. The blank, HnD and OBnD phantoms were triggered with 10.5 MPa PNP (I\textsubscript{SPPA} = $3057 W/cm^2$) while the PnDs were triggered with 8.4 MPa PNP(I\textsubscript{SPPA} = $1956 W/cm^2$). The PNPs chosen for this experiment were chosen based on similar literature and experimentation for ADV of HnDs\cite{Burgess2019ControlNanoemulsionsb,Shpak2014AcousticFocusing.}. To ensure a fair comparison with PnDs a lower PNP was used to prevent possible excess cavitation of the PnDs. While there was no substantial difference in findings between 8.4 and 10.5 MPa for the PnDs, the lower PNP results are reported here to ensure a fair comparison between PnDs and HnDs. Each HIFU pulse was separated from the next by approximately 5 ms. 

For the second set of experiments 21 HnD phantoms were prepared. These phantoms were imaged at 0 \degree C, 23 \degree C, or 37 \degree C to investigate the relationship between the ADV threshold and temperature. The temperature was controlled by acclimatizing the phantoms for 20 minutes in either an ice bath or a 37 \degree C heated bath with closed-loop control using a probe thermometer. The PA phantoms were then removed for less than 30 seconds while a single image set was acquired and returned to their respective bath for one minute before the next image set. The change in phantom temperature over these 30 seconds was measured to be 0.36 $\pm$ 0.11 degrees C. Seven phantoms were tested for each temperature, with each pressure level applied to a unique spot on the phantom. Each image acquisition consisted of three HIFU activations per spot for a total of 21 vaporization measurements per sample. The comparison of ultrasound image signal in the HIFU focus before and after the HIFU pulse was taken to be a metric for vaporization. The threshold for vaporization was defined as a five percent increase in ultrasound signal in the focal area.
\begin{figure*}[b]
\includegraphics[scale=.3]{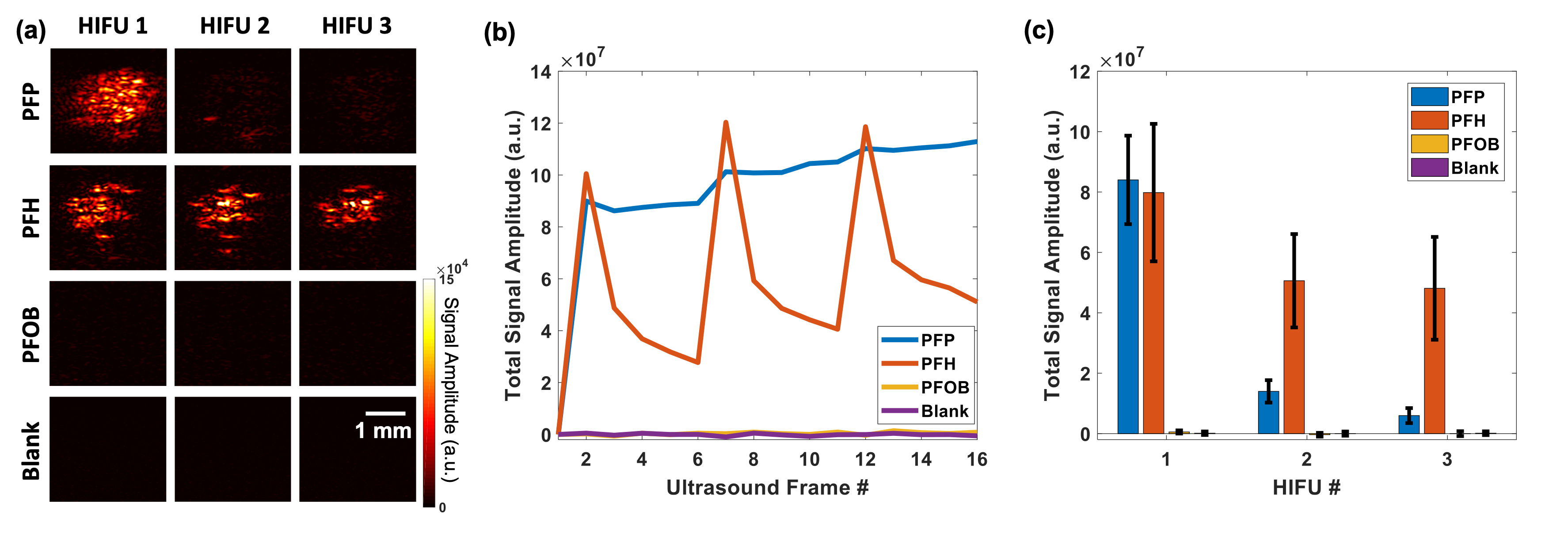}
\centering

\caption{Comparison of PFCnDs in Polyacrylamide phantoms. (a) Example differential images of three consecutive HIFU activations are shown for phantoms containing the three types of perfluorocarbon nanodroplets and one blank control phantom. (b) Total signal amplitude in the focal area is imaged over three consecutive vaporization events. Example signature time traces show nanodroplet behavior for different perfluorocarbon nanodroplets. (c) Total signal amplitude in the focal area is averaged over 5 phantoms to show the sustained vaporization of HnDs.}
\label{fig:p1}
\end{figure*}

For the final set of experiments three HnD phantoms were prepared. These phantoms at room temperature were repeatedly vaporized with 10.5-MPa HIFU with a pulse repetition frequency of 10 Hz. The HIFU was applied to the same spot to investigate the robustness of HnDs over the course of multiple vaporization/recondensation cycles. The phantom was imaged at defined intervals after 0, 100, 750, 1,500, and 10,000 HIFU pulses.

\section{Results}
The PFCnDs exhibited similar sub-micron size distributions. The averaged peak diameter for the PnD samples was 492.24 $\pm$ 39.85 nm at an average concentration of $2.4*10 ^8$ particles/mL. The averaged peak diameter for the HnD samples was 558.71 $\pm$ 31.58 nm at an average concentration of $1.8*10 ^8$  particles/mL. The averaged peak diameter for the OBnD samples was 455.93 $\pm$ 58.25 nm at an average concentration of $2.9*10 ^8$ particles/mL.

To demonstrate that HnDs can be repeatedly vaporized with HIFU, we first confirmed that ultrasound signal in the HIFU focus requires the presence of activatable PFCnDs and that the repeatable vaporization is due to the high boiling point perfluorocarbon core. To this end, PA phantoms were prepared containing no nanodroplets, PnDs, HnDs or OBnDs. Blank polyacrylamide phantoms served as a control to test for the possibility that the visible time-varying ultrasound signal was due to cavitation in the phantom. OBnDs, with a core perfluorocarbon boiling point of 142 \degree C, served as a secondary control to test whether cavitation was occurring due to the nanodroplets acting as nucleation sites. 

Figure \ref{fig:p1}(a) shows examples of differential ultrasound images acquired with three consecutive HIFU pulses for each of the four phantom groups. It is apparent there was no significant ultrasound signal increase in the blank or OBnD phantoms for any of the HIFU pulses.  In the phantoms with PnDs, the first HIFU pulse triggered many of the nanodroplets resulting in the visible echogenic cloud. Subsequent HIFU pulses failed to produce more microbubbles as most nanodroplets in the focal spot were already vaporized. Single-shot PnD activity is supported by previous studies\cite{Ishijima2016TheNano-droplets,Rojas2019VaporizationRatio,Rapoport2009ControlledNanoemulsions/microbubbles,Reznik2013TheDroplets}. The representative HnD image set demonstrates the repeatable acoustic vaporization of HnDs compared to PnDs. Each HIFU pulse produced a similar vaporization response from the HnDs. 

Figure \ref{fig:p1}(b) and \ref{fig:p1}(c) further illustrate the principal results of these images. Figure \ref{fig:p1}(b) tracks the total sum of ultrasound signal amplitude in the focal area over a three HIFU vaporization events in a 16-frame ultrasound imaging sequence. The HIFU pulses occur prior to ultrasound frames 2, 7, and 12. PnDs vaporize on the first HIFU pulse with marginal increases on the second and third pulses. HnDs vaporize after each HIFU pulse and the nanodroplets quickly recondense resulting in decaying ultrasound signal after each peak. Figure \ref{fig:p1}(c) shows the increase in ultrasound signal magnitude after three HIFU stimuli averaged over five phantoms. Taken together, these results show that HnDs can be reliably and repeatedly vaporized with a HIFU stimulus.

\begin{figure}[h]
\includegraphics[scale=.3]{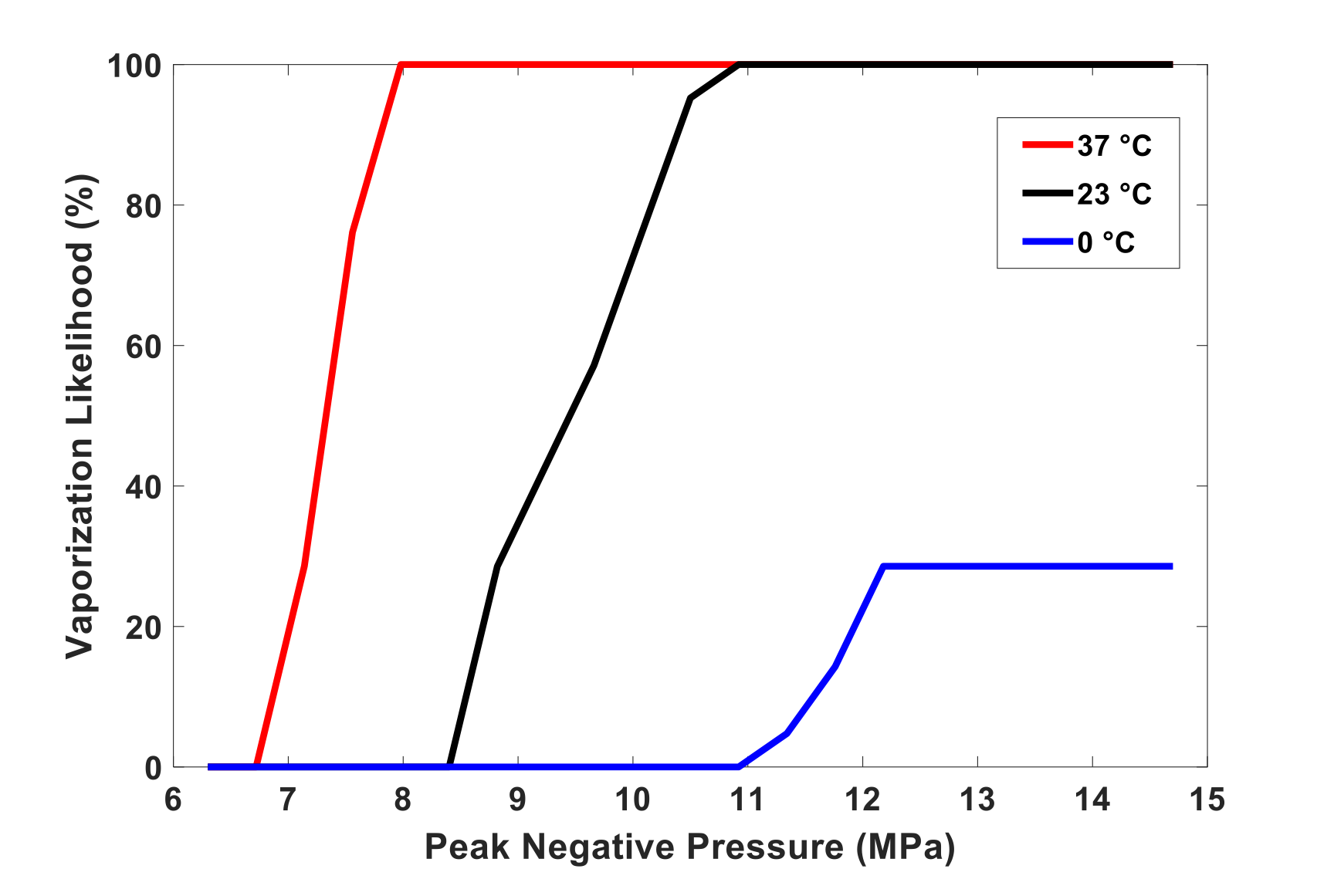}
\centering

\caption{A visualization of the percent chance for vaporization for three temperatures with increasing HIFU peak negative pressure. The vaporization likelihood is defined as a threshold 5 percent increase in total ultrasound signal amplitude in the focal area post-HIFU. Seven phantoms were tested for each temperature with 3 HIFU pulses per tested peak negative pressure.}
\label{fig:thresh}
\end{figure}

The thresholds for vaporization of HnDs were further examined in Figure \ref{fig:thresh}. A sufficiently large PNP from the HIFU is required to overcome the vaporization pressure of the high boiling point of perfluorohexane. HnD phantoms were subjected to HIFU pulses with increasing amplitude at 0 \degree C, 23 \degree C, and 37 \degree C. Figure \ref{fig:thresh} shows a decreasing vaporization threshold with increasing temperature. A threshold for reliable activation at room temperature is 10.5 MPa. This acoustic threshold varies with temperature because environmental temperature affects the pressure required to instigate the phase change and overcome the perfluorocarbon boiling point and stabilizing Laplace pressure. The temperature dependence of ADV lowers the activation threshold to 8 MPa at body temperature, with some phantoms showing vaporization with a PNP as low as 7.1 MPa. Therefore, a transition to \textit{in-vivo} work would require a smaller peak negative pressure than the other studies conducted in this work, which were performed at room temperature. At 0 \degree C, the HnDs showed no vaporization activity at less than 11.3 MPa. At PNPs of 11.3 MPa and greater, two of the seven low-temperature phantoms clearly exhibited vaporization onward while five showed no activity.

\begin{figure}[h]
\includegraphics[scale=.28]{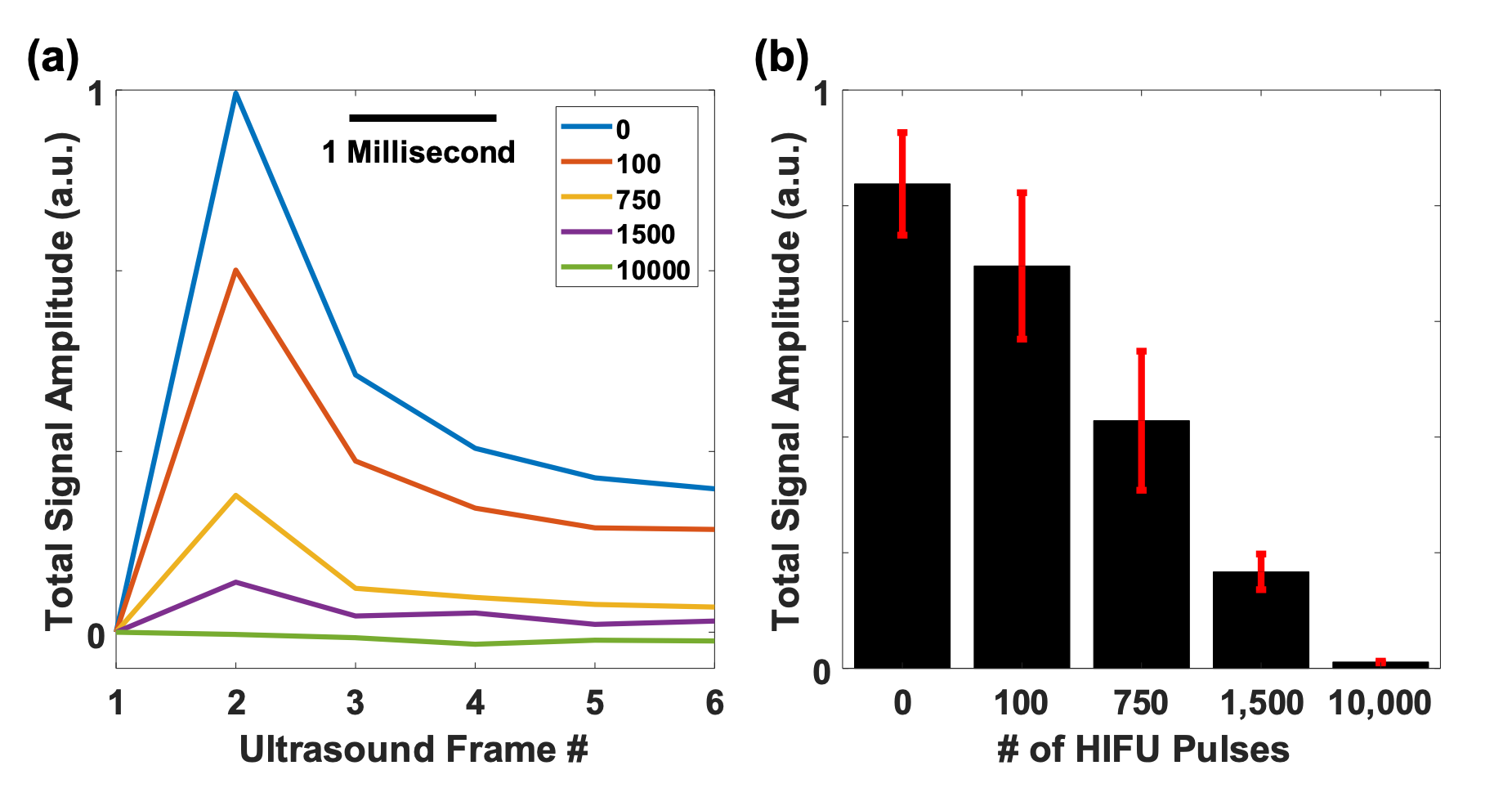}
\centering

\caption{Lifetime of repeatable recondensation. (a) Summation of ultrasound signal in the focal spot traced over time taken after X (as indicated in the legend) number of previous HIFU pulses. (b) Averaged total signal across three trials shows consistent vaporization during for the first few hundred vaporization events, but diminishing signal after thousands of HIFU pulses as fewer HnDs are vaporizing.}
\label{fig:cycle}
\end{figure}

The primary motivation for using a high boiling point perfluorocarbon is the prospect of sustained imaging of PFCnDs.  Thus, we investigated the number of HnD vaporization/recondensation cycles which could be reasonably imaged. This experiment was done at room temperature to ensure temperature stability of the phantom over the long duration of this experiment. Based on the results from Figure \ref{fig:thresh}, a peak negative pressure of 10.5 MPa was used to ensure vaporization of HnDs. The HIFU focus was kept static so as to repeatedly vaporize the same population of HnDs. The phantoms were exposed to 0, 100, 750, 1,500, or 10,000 HIFU pulses prior to undergoing ultrasound visualization. Figure \ref{fig:cycle}(a) shows representative examples of vaporization and recondensation after each number of HIFU pulses to test the diminishing vaporization response and durability of the nanodroplets. The signal amplitude was normalized to the pre-HIFU level. The characteristic exponential decay of ultrasound signal in frames 2-6 is due to the stochastic recondensation of the HnDs after the HIFU-induced vaporization. Figure \ref{fig:cycle}(b) shows the averaged signal amplitude of this repeated vaporization experiment from three phantoms. The HnD signal amplitude gradually decreases over the course of hundreds to thousands of vaporization events. Even after the 10,000 HIFU pulses there was no visible damage to the phantom. These data show that the HnDs remain remain highly stable even after minutes of sustained HIFU-induced vaporization.

\section{Discussion}

High boiling point nanodroplets such as HnDs offer the benefit of repeatable vaporization. When applied to \textit{in-vivo} imaging, the HnDs will be less volatile than their lower boiling point counterparts. In addition, each nanodroplet can be activated and imaged multiple times. This repeated vaporization can be used to suppress the background tissue and boost the sensitivity of the method\cite{Hannah2016BlinkingImaging.}. The dynamic contrast could also be leveraged to employ super-resolution imaging methods\cite{Luke2016Super-ResolutionNanodroplets}. Importantly, this could be applied to stationary HnDs, rather than relying on flowing microbubbles\cite{Errico2015UltrafastImaging}. This has the potential to enable high-resolution molecular imaging.

The benefits of repeatable activation come with the drawback of a higher activation threshold. Some groups have shown this threshold can be decreased through innovative strategies such as incorporating nucleation sites and multi-modal activation \cite{Li2020SpatiallyAgents,Strohm2011VaporizationIrradiation,Vezeridis2019Fluorous-phasePoint}.In this study, activation is achieved with high-amplitude, short pulses of HIFU. This waveform has a similar pulse duration to mechanical tissue disruption techniques, such as lithotripsy and histotripsy\cite{Ikeda2016FocusedLithotripsy}. The short pulse duration employed by these methods results in a negligible temperature rise in the tissue\cite{Khokhlova2015HistotripsyApplications,Hoogenboom2015MechanicalEffects}. Given that the current work uses much lower pressure levels, it is unlikely that thermal tissue damage will occur. Despite the comparatively lower HIFU energies, the potential for mechanical damage via cavitation still exists. As shown in Figure \ref{fig:thresh}, the minimum activation peak negative pressure at body temperature for 100\% chance of activation is 8 MPa, which corresponds to a mechanical index of 7.6. This MPa is well below reported thresholds of 13.7 to 21 MPa for 1.1-MHz ultrasound  cavitation in tissue\cite{Maxwell2013ProbabilityMaterials.}, but the mechanical index is above the accepted diagnostic ultrasound imaging threshold of 1.9\cite{Sen2015MechanicalIndex}. While the cavitation threshold depends on experimental variables such as dissolved gas content, nucleation sites, and water concentration\cite{Rojas2019VaporizationRatio,Raut2019ImpactUltrasound,Burgess2019ControlNanoemulsionsb,Reznik2013TheDroplets,Xu2018AcousticUltrasound}, we found the energy levels required for HnD vaporization to be lower than those of cavitation in blank phantoms or OBnD phantoms with nanodroplet nucleation sites. It is important to note this threshold is also dependent on the imaging environment. Phantom damage was only observed in the case of peak negative pressures greater than 12.2 MPa in HnD phantoms at 37 \degree C. We hypothesize that this could be the result of a more violent vaporization process, or ultrasonic cavitation after the initial formation of microbubbles. Further \textit{in-vivo} testing of the mechanical effects is needed.

The acoustic vaporization threshold for phase change particles varies greatly in literature \cite{Lea-Banks2019Ultrasound-responsiveReview,Giesecke2003Ultrasound-mediatedVitro,Dayton2006ApplicationTherapy,Sheeran2016Phase-ChangeTherapy,Loskutova2019ReviewTherapeutics,Xu2018AcousticUltrasound,Aliabouzar2019EffectsThreshold}. The perfluorocarbon core boiling point is one clear factor, but the Laplace pressure, shell composition, and phantom stiffness also play a role in this threshold. While PFCnDs can be synthesized with diameters of smaller than 200 nm, using larger nanodroplets could decrease the activation energy threshold\cite{Kripfgans2004OnDroplets}. This is due to a lower Laplace pressure counteracting against vaporization\cite{Rapoport2009ControlledNanoemulsions/microbubbles,Sheeran2011FormulationUltrasound}. While not fully understood, it is likely that the presence of a nucleation event is required to initiate the gaseous phase and therefore sufficient ultrasound does not ensure this vaporization mechanism\cite{Shpak2014AcousticFocusing.}. It is also possible that microbubbles could provide nucleation sites and greater energy absorption, thereby decreasing the energy needed for subsequent HIFU pulses. An interesting observation is the decreased signal from the HnDs after many vaporization cycles. We hypothesize that with each activation of the HnDs there is a chance that an individual microbubble is destroyed by cavitation or coalesces with neighboring microbubbles and no longer responds to ultrasound activation. There is also the possibility that a small amount of perfluorocarbon escapes with each microbubble transition cycle. The mechanisms of degradation requires further study. 

 While the focus here is on acoustic activation of HnDs, similar PFCnDs have been constructed for repeated optical activation with a pulsed laser\cite{Hannah2016BlinkingImaging.,Luke2016Super-ResolutionNanodroplets,Asami2012RepeatableImaging}. Optical activation of PFCnDs offers unique benefits, including multiplex and multimodal imaging\cite{Lin2017OpticallyImaging,Santiesteban2019Color-codedImaging}. This comes with the limitation that optical energy is quickly attenuated in tissue via scattering and absorption, making deep activation difficult. With human skin thickness varying between 1-4 mm, optical activation of HnDs works well for small animal models but would be restricted to surface or near-surface clinical applications. For applications deeper in tissue than 1 cm, ultrasound travels through tissue with minimal attenuation. Although the PA transduction cone is less acoustically scattering than tissue, the HIFU wave in this study traveled 9 cm before reaching the PFCnD samples. One limitation with the experimental setup is the inability to utilize the full extent of acoustic activation's enhanced penetration depth. The right angle geometry of the HIFU and imaging transducer shown in Figure \ref{fig:setup} is infeasible for deep \textit{in-vivo} experiments. We envision an annular HIFU array with a coaxial imaging transducer would enable deeper imaging in tissue. Alternatively, acoustic activation has a better depth of penetration and pulse repetition frequency. While the type of activation energy would probably be chosen based on the application, it is possible to image the same particles using both methods\cite{Li2020SpatiallyAgents}.

\section{Conclusion}
Phase change nanodroplets have great potential as multifunctional,  nano-sized ultrasound contrast agents. Their unique kinetics can be visualized in real-time, noninvasively, with a non-ionizing imaging modality.  We systematically studied the acoustic droplet vaporization of high boiling point perfluorocarbon nanodroplets and identified their ability to recondense. Benefits of acoustic activation were highlighted to bring perfluorohexane phase change nanodroplet detection to clinically relevant depths.  The ADV threshold was shown to lower with increasing temperature, lowering the energy requirements to extend these particles \textit{in-vivo}. Rapid, repeated vaporization of these particles provides a platform for clinically relevant tracking of blinking particles and their kinetics.
 
\section*{Acknowledgment}

The authors would like to thank Prof. Tyrone Porter from the University of Texas for the needle hydrophone measurements of our experimental setup.  Support was provided by NIH grant  R21CA197409 and by a Prouty Pilot Grant from the Norris Cotton Cancer Center.





\bibliographystyle{IEEEtran}
\bibliography{references.bib}

\begin{IEEEbiography}[{\includegraphics[width=1in,height=1.25in,clip,keepaspectratio]{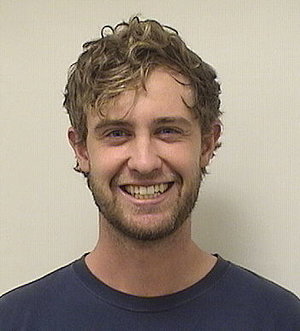}}]{Austin Van Namen}
Austin Van Namen is a Ph.D. student in the Thayer School of Engineering at Dartmouth College. He earned his B.E. in Biomedical Engineering and Mathematics from Vanderbilt University. The focus of his research bridges nanoscience, mathematics and imaging science. His current research projects revolve around combining multi-functional nanoparticle contrast agents with ultrasound and photoacoustic imaging.
\end{IEEEbiography}
\vskip 0pt plus -1fil
\begin{IEEEbiography}[{\includegraphics[width=1in,height=1.25in,clip,keepaspectratio]{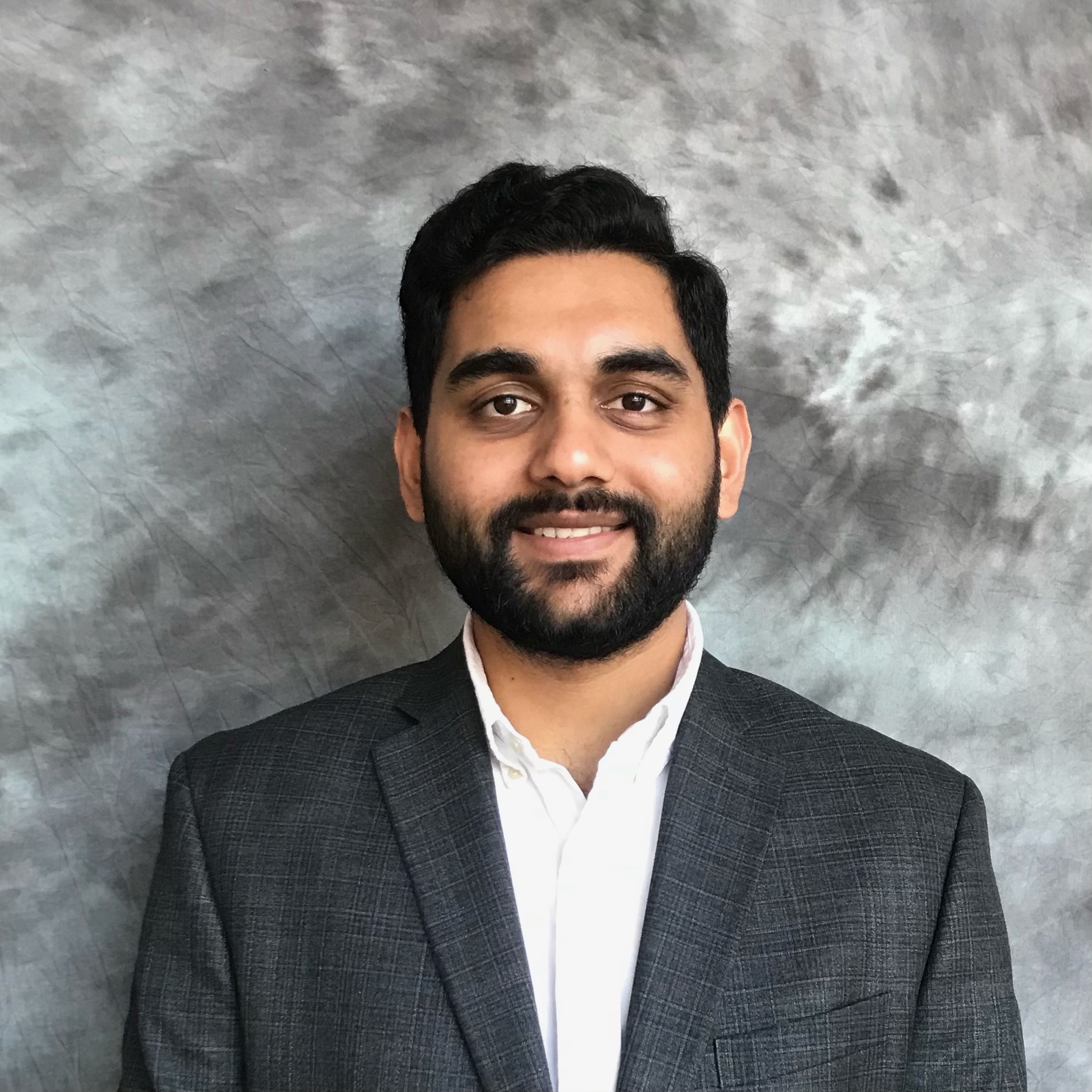}}]{Sidhartha Jandhyala}
Sidhartha Jandhyala is a Ph.D. Student at the Thayer School of Engineering. He received his B.S in Biomedical Engineering from North Carolina State University. His current research project is focused on combining nanoparticle contrast agents with ultrasound and photoacoustic imaging for delivery and monitoring of oxygen to tumors. 
\end{IEEEbiography}
\vskip 0pt plus -1fil

\begin{IEEEbiography}[{\includegraphics[width=1in,height=1.25in,clip,keepaspectratio]{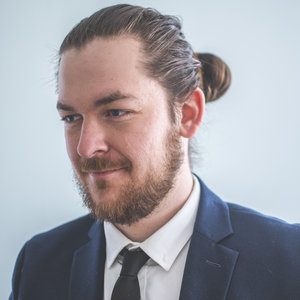}}]{Tomas Jordan}
Tomas Jordan is a Ph.D. student in the Thayer School of Engineering at Dartmouth College. He received his B.S. in Biomedical Engineering from Boston University. He is currently developing a method for non-invasive stimulation and imaging of neural activity using piezoelectric nanoparticles and ultrasound. His other research interests are peripheral nerve repair and applications of sonoluminiscence in medical imaging. 
\end{IEEEbiography}
\vskip 0pt plus -1fil

\begin{IEEEbiography}[{\includegraphics[width=1in,height=1.25in,clip,keepaspectratio]{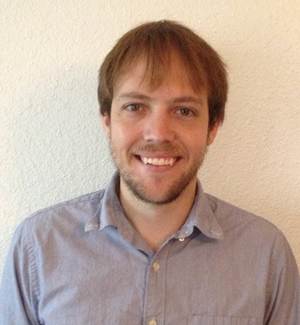}}]{Geoffrey Luke}
Geoffrey P. Luke, PhD is an assistant professor in the Thayer School of Engineering at Dartmouth College and a member of the Cancer Imaging and Radiobiology Research Program at Norris Cotton Cancer Center. He directs the Functional and Molecular Imaging Research Laboratory. He earned his B.S. in Computer Engineering and Mathematics as well as his M.S. in Electrical Engineering from the University of Wyoming. Dr. Luke completed his Ph.D. in Electrical Engineering at The University of Texas at Austin.
\end{IEEEbiography}
\vskip 0pt plus -1fil

\end{document}